\documentclass[aps,showpacs,twocolumn]{revtex4}
\usepackage[dvips]{graphicx}
\usepackage{amssymb}
\newcommand{\ped}[1]{\ensuremath{_{\rm #1}}}

\begin{document}

\title{Two-band Eliashberg equations and the experimental $T_{c}$
of the diboride $Mg_{1-x}Al_{x}B_{2}$}

\author{G.A. Ummarino, R.S. Gonnelli}
\affiliation{$INFM-$Dipartimento di Fisica, Politecnico di Torino,
Corso Duca degli Abruzzi 24, 10129 Torino, Italy\\ E-mail address:
giovanni.ummarino@infm.polito.it}
\author{S. Massidda}
\affiliation{$INFM-$Dipartimento di Fisica, Universit$\grave{a}$
di Cagliari, Cittadella Universitaria SP Monserrato-Sestu, km
0,700 I-09042 Monserrato, Italy}
\author{A. Bianconi}
\affiliation{$INFM$ - Dipartimento di Fisica, Universit$\grave{a}$
di Roma
"La Sapienza"\\
 Piazzale Aldo Moro 2, 00185 Roma, Italy}
\begin{abstract}
The variation of the superconducting critical temperature $T_{c}$
as a function of $x$ in the diboride $Mg_{1-x}Al_{x}B_{2}$ has
been studied in the framework of the two-bands Eliashberg theory
and traditional phonon coupling mechanism. We have solved the
two-bands Eliashberg equations using first-principle calculations
or simple assumptions for the variation of the relevant physical
quantities. We have found that the experimental $T_{c}$ curve can
be explained only if the Coulomb pseudopotential changes with $x$
by tuning  the Fermi level toward the sigma band edge. In
polycrystal samples the $x$ dependence of the $\sigma$ and
$\pi$-band gap has been found and is in agreement with
experiments.
\end{abstract}


\maketitle

The recent discovery of superconductivity at T=40 K in magnesium
diboride \cite{refe1t} has stimulated intense investigation, both
from the theoretical and the experimental point of view. Now the
electronic structure of $MgB_{2}$ is well understood and the Fermi
surface consists of two three-dimensional sheets, from the $\pi$
bonding and antibonding bands, and two nearly cylindrical sheets
from the two-dimensional $\sigma$ bands \cite{refe2t}. There is a
large difference in the electron-phonon coupling on different
Fermi surface sheets and this fact leads to a multiband
description of superconductivity. The superconductivity in
$MgB_{2}$ has been deeply studied in the past two years. More
recently the substitutions of $Mg$ with aluminum \cite{refe2at}
have been investigated to understand the evolution of the pairing
process by tuning the Fermi level toward the top of the sigma band
edge. The $Mg_{1-x}Al_{x}B_{2}$ alloys show
\cite{refe3t,refe4t,refe4at} a continuous evolution through a
complicated mixed phase from $MgB_{2}$ (x=0) to the final member
$AlMgB_{4}$ (x=0.5) where an ordered superlattice structure of
boron layers intercalated by alternating layers of $Al$ and $Mg$
is formed. Even though the alloys with intermediate $x$ are rather
disordered, their $T_{c}$ is well defined and drops with
decreasing $x$,  becoming zero \cite{refe3t,refe4t,refe4at} for
$x>0.5$. We can see in Fig. 1 the experimental data of critical
temperature taken from ref. 4-6. Two ranges of variation are
present: a "high-$T_{c}$" range for $0<x<0.3$ that shows a slow
variation with x, while a rapid variation occurs in the
"low-$T_{c}$" range $0.3<x<0.5$. This variation has been
interpreted as to a 2D-3D cross-over, at $x\cong 0.33$, of the
topology of the Fermi surface \cite{refe4t,refe4at}.
Theoretically, the transition temperature $T_{c}$ in
$Al$-substituted $MgB_{2}$ has been studied \cite{refe5at}
 within the two-band BCS formalism, as a function of
the aluminum content; the variation of the effective interband
coupling as a function of $x$ has been obtained from the
experimental values of $T_{c}(x)$. Ab-initio calculations have
been performed on $Al$-substituted $MgB_{2}$ by Profeta {\em et
al.}\cite{Mass} and by de la Pe${\tilde n}$a {\em et
al.}\cite{delapena}, within the virtual crystal approximation.
These works correctly describe the experimental trends; in
particular\cite{Mass}, the behavior of the $E_{2g}$ phonon
frequency and of the critical temperature agree reasonably well
with experiment.

\begin{figure}[!]
 \begin{center}
 \includegraphics[keepaspectratio, width=0.9\columnwidth]{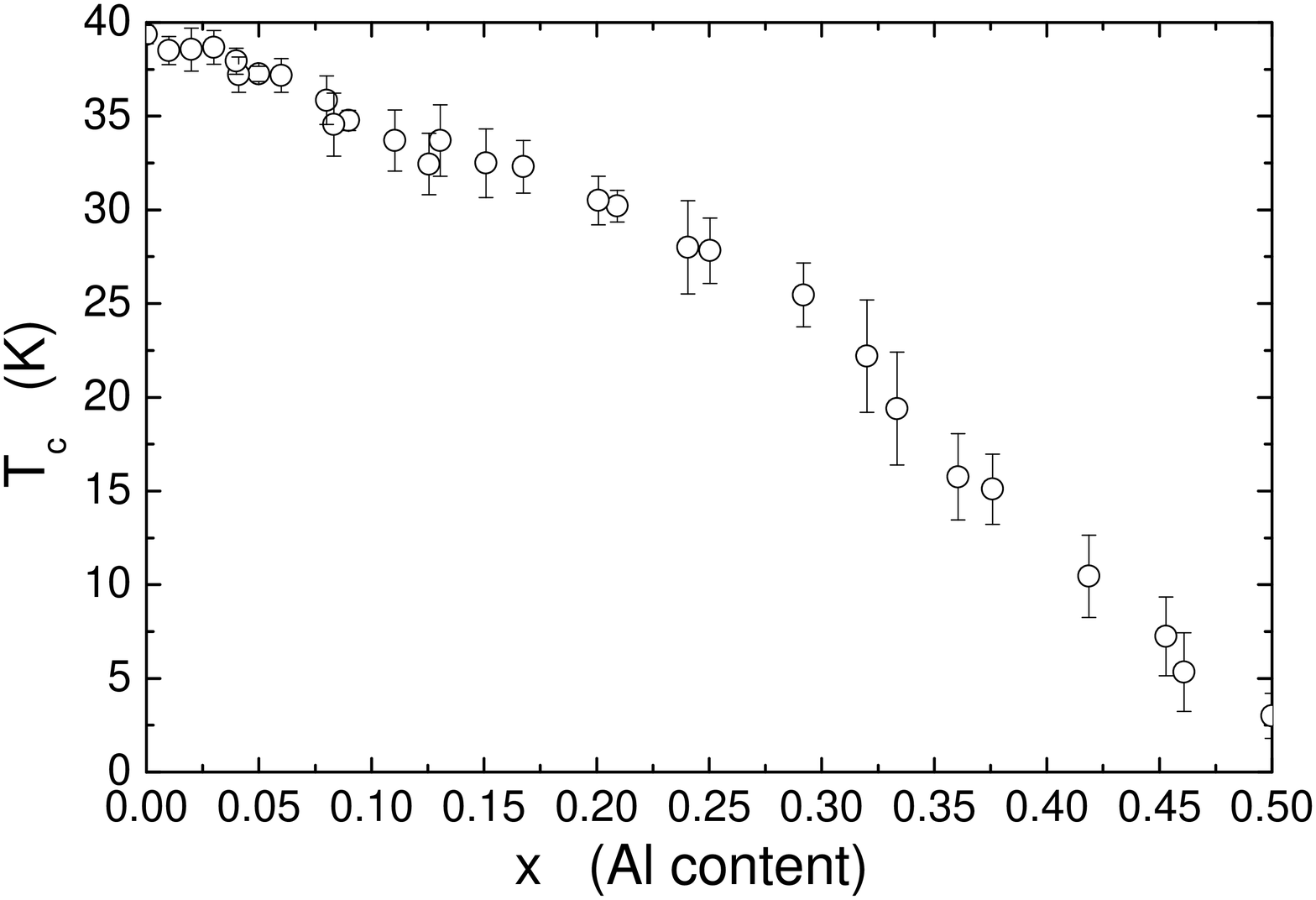}
 \end{center}
 \caption{The value of the experimental $T_{c}$ (open circle)
 as a function of x, taken from ref. 4-6.}
 \label{fig:1}
 \end{figure}

In this work we further investigate $Mg_{1-x}Al_{x}B_{2}$ (for
$0\le x \le 0.5$) by computing the variation of the critical
temperature as a function of $x$ by means of the Migdal-Eliashberg
two bands theory. For the peculiar characteristics of our system
we must use a generalization of the Eliashberg theory
\cite{refe5t,refe6t,refe7t} for systems with two bands
\cite{refe10t,refe11t}. The two bands Eliashberg equations have
already been used for studying the $MgB_{2}$ system
\cite{refe14t,refenoi,refe16at,refe16t}. Here we have eight input
parameters: four (but only three independent \cite{refe11t})
electron phonon spectral functions
$\alpha^{2}_{ij}(\omega)F_{ij}(\omega)$ and four (but only three
independent) Coulomb pseudopotential $\mu^{*}_{ij}(\omega_{c})$ to
reproduce the experimental values of $T_\mathrm{c}$ versus $x$ by
solving the Eliashberg equations. In this case there are four
equations to be solved for the calculation of the gaps
$\Delta_{i}(i\omega_{n})$ and the renormalization functions
$Z_{i}(i\omega_{n})$.
 In the approximation of a flat normal densities of states and infinite bands,
 the s-wave two bands Eliashberg equations are:
 \begin{eqnarray}
\omega_{n}Z_{i}(i\omega_{n})&=&\omega_{n}+\pi
T\sum_{m,j}\Lambda_{ij}(i\omega_{n}-i\omega_{m})N^{j}_{Z}(i\omega_{m})+\nonumber\\
& & +\sum_{j}\Gamma^{ij}N^{j}_{Z}(i\omega_{n})\label{eq:EE1}
\end{eqnarray}
\begin{eqnarray}
Z_{i}(i\omega_{n})\Delta_{i}(i\omega_{n})&=&\pi
T\sum_{m,j}[\Lambda_{ij}(i\omega_{n}-i\omega_{m})-\mu^{*}_{ij}(\omega_{c})]\cdot\nonumber\\
& &
\hspace{-1.5cm}\cdot\theta(|\omega_{c}|-\omega_{m})N^{j}_{\Delta}(i\omega_{m})+\sum_{j}%
\Gamma^{ij}N^{j}_{\Delta}(i\omega_{n}) \label{eq:EE2}
\end{eqnarray}
 where $\theta$ is the Heaviside function, $\omega_{c}$ is a
 cut-off energy, $\Gamma^{ij}$ is the non magnetic impurity scattering rate in the Born
 approximation and
\begin{equation}
\Lambda_{ij}(i\omega_{n}-i\omega_{m})=\int_{0}^{+\infty}\frac{d\omega
\alpha^{2}_{ij}F(\omega)}{(\omega_{n}-\omega_{m})^{2}+\omega^{2}}
\end{equation}
\begin{equation}
N^{j}_{\Delta}(i\omega_{m})=\frac{\Delta_{j}(i\omega_{m})Z_{j}(i\omega_{m})}{\sqrt{\omega^{2}_{m}Z_{j}^{2}(i\omega_{m})+\Delta^{2}_{j}(i\omega_{m})Z_{j}^{2}(i\omega_{m})}}
\end{equation}
\begin{equation}
N^{j}_{Z}(i\omega_{m})=\frac{\omega_{m}Z_{j}(i\omega_{m})}{\sqrt{\omega^{2}_{m}Z_{j}^{2}(i\omega_{m})+\Delta^{2}_{j}(i\omega_{m})Z_{j}^{2}(i\omega_{m})}}
\end{equation}
%
where $\omega_{n}=\pi T(2n-1)$ and $n, m=0,\pm 1,\pm 2...$. In
writing the Eliashberg equations we have neglected the asymmetric
part of the self-energy $\chi_{i}(i\omega_{n})$ (which is always
equal to zero in the half-filling case) and the equation that
represents the conservation of the particles necessary for
calculating the shift of the chemical potential. These assumptions
are correct if the normal density of states is symmetric or
constant as function of energy and the bands are infinite. We have
calculated the normal densities of state of the $\sigma$ and $\pi$
band and, with good approximation, they are constant around the
Fermi energy for all $x$-values. Finally to use the infinite band
approximation is correct for the $\pi$-band (half-bandwidth
$W_{\pi}(x=0)\simeq$ 3 eV) while we can have some doubt for the
$\sigma$-band ($W_{\sigma}(x=0)\simeq$ 0.5 eV)\cite{emamgb2}. If
we solve the Eliashberg equations with these last finite values of
the bands \cite{refe7t,band} for the $MgB_{2}$ ($x=0$ case), by
using the same parameters of the infinite bands approximation, we
find $T_{c}$ =40 K. For having $T_{c}$ =39.4 K it must be
$\mu(\omega_{c})$=0.0333 (see eq. 10) instead of $0.0326$. Since
in the first case $T_{c}$ is only slightly lower than the
experimental one and $\mu(\omega_{c})$, in the second case, is
almost the same as in the infinite band case so we can conclude
that to assume the bands as infinite is a plausible approximation.
The situation can be different, of course, for large values of
$x$.
\begin{figure}[!]
 \begin{center}
 \includegraphics[keepaspectratio, width=0.9\columnwidth]{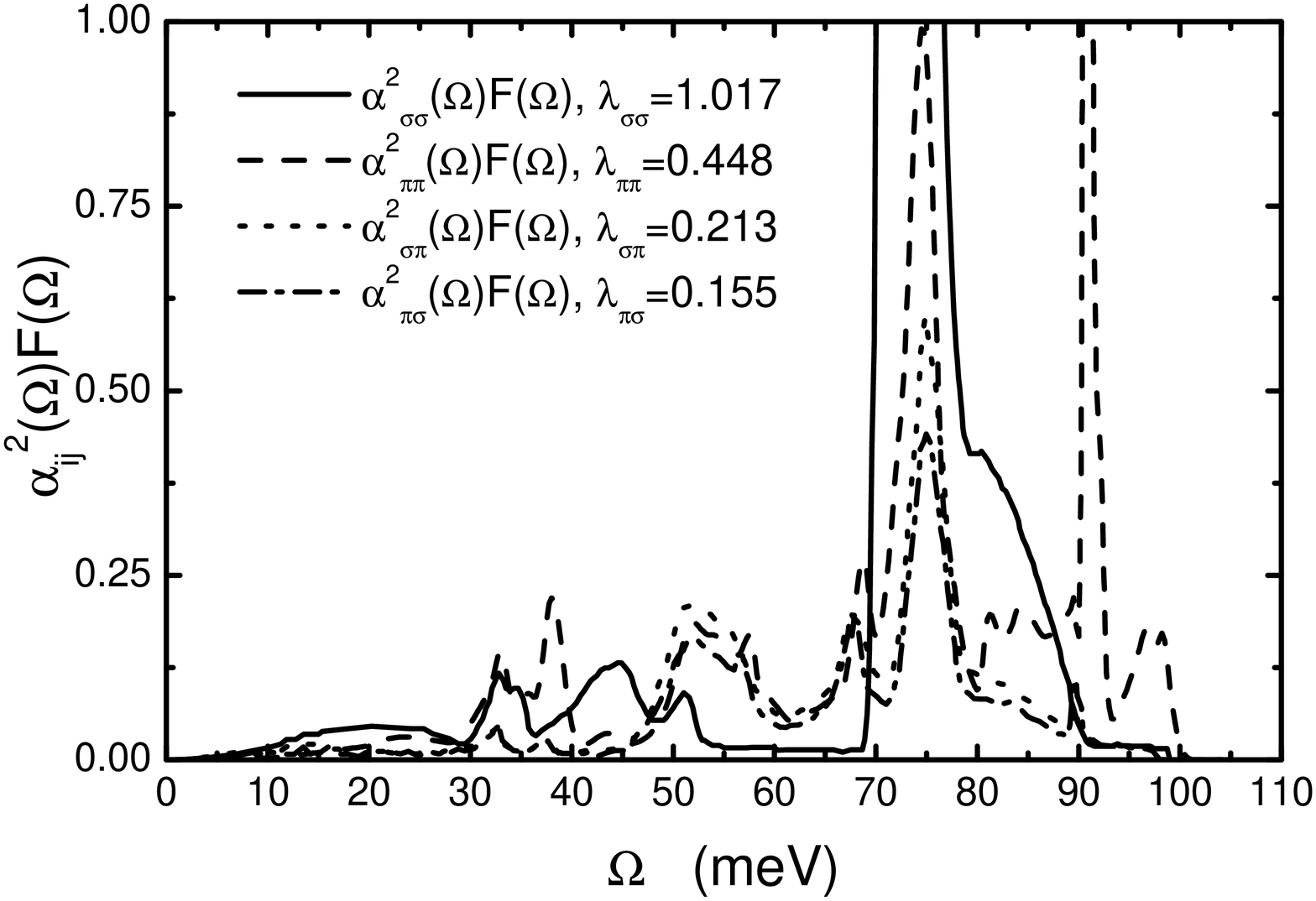}
 \end{center}
 \caption{The spectral functions of the two-band model for the $MgB_{2}$:
$\sigma\sigma$ (solid line), $\pi\pi$ (dashed line), $\sigma\pi$
(dotted line) and $\pi\sigma$ (dashed dotted line), taken from
ref. 15.} \label{fig:2}
 \end{figure}
\begin{figure}[t]
 \begin{center}
 \includegraphics[keepaspectratio, width=0.9\columnwidth]{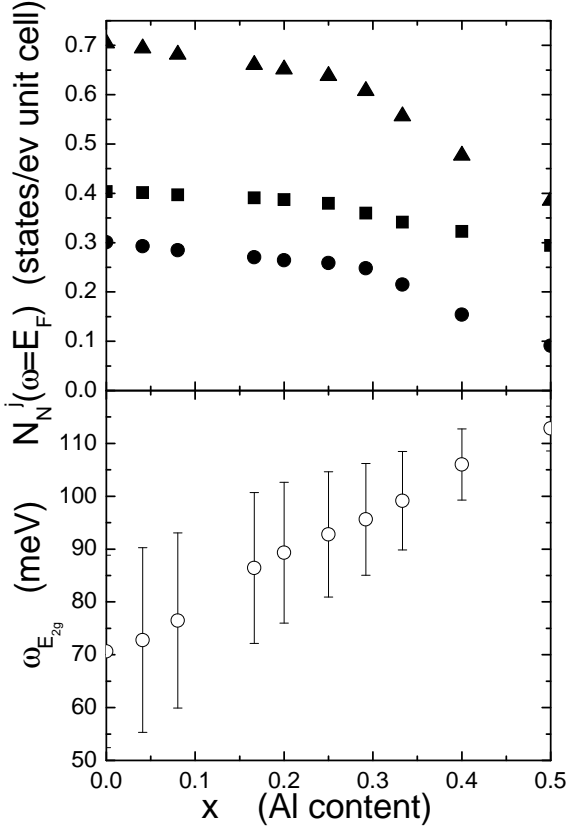}
 \end{center}
 \caption{Upper panel: calculated density of
states at the Fermi energy $N_{N}(\omega=E_{F},x)$ in the
$\sigma$-band (filled circles), in the $\pi$-band (filled squares)
and total (filled up triangles) as a functions of $x$. Lower
panel: the phonon  frequency $\omega_{E_{2g}}$ obtained from
experimental data as a function of $x$, see ref. 6.} \label{fig:3}
 \end{figure}
\begin{figure}[t]
 \begin{center}
 \includegraphics[keepaspectratio, width=\columnwidth]{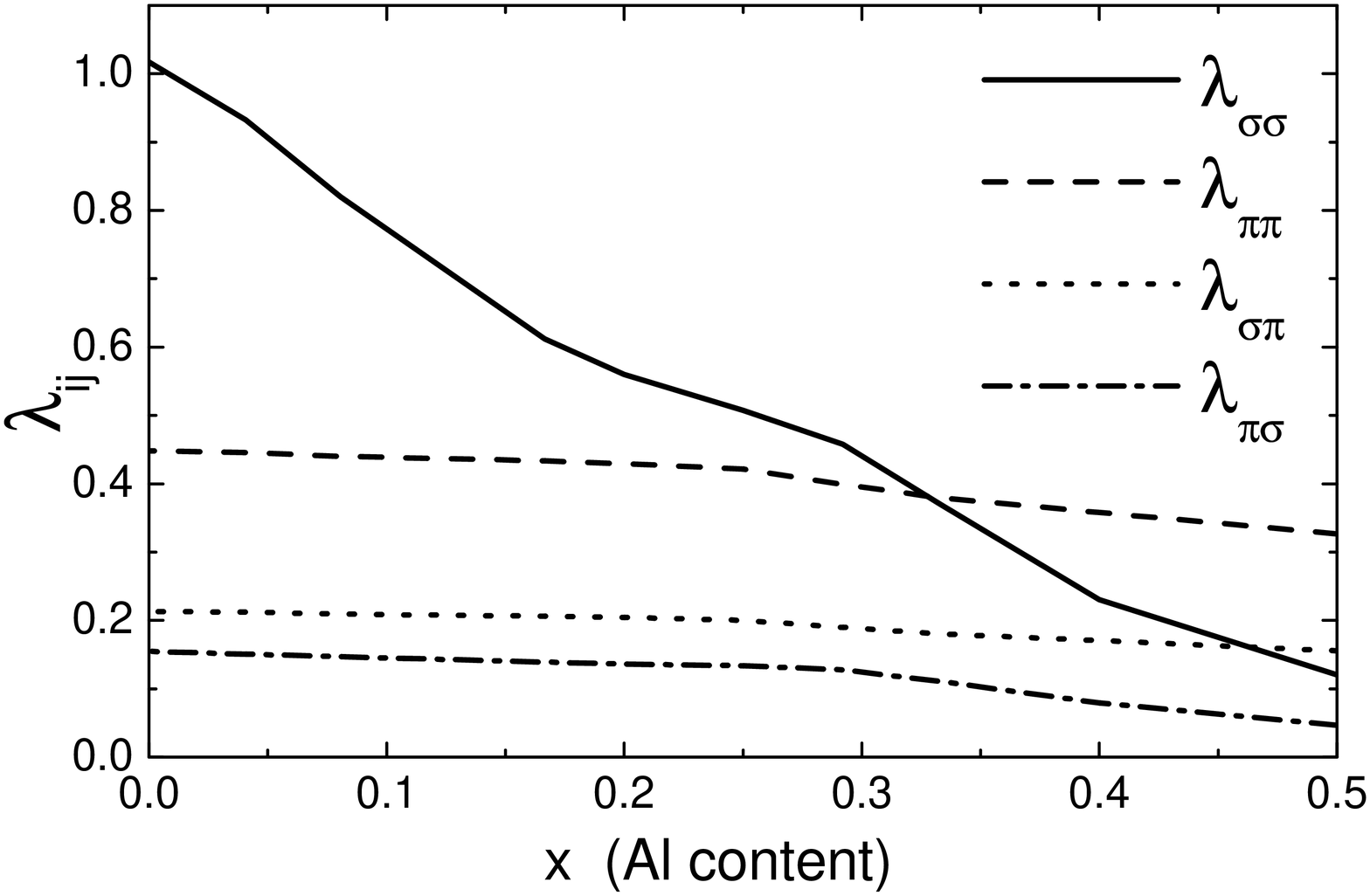}
 \end{center}
 \vspace{8mm} \caption{Calculated electron phonon coupling constants
$\lambda_{ij}$ as a function of $x$ in the four cases:
$\sigma\sigma$ (solid line), $\pi\pi$ (dashed line), $\sigma\pi$
(dotted line) and $\pi\sigma$ (dashed dotted line).} \label{fig:4}
 \end{figure}

Finally, one can think that it would be simpler to use the Suhl
formula \cite{Suhl} for explaining the $T_{c}$ versus $x$ curve
without solving the complete two-band Eliashberg equations. The
problem of the Suhl formula is the prefactor before the
exponential: it is a phononic energy but in a two-band theory it
is difficult to specify it exactly and so it is considered as
another free parameter.

For solving the two-band Eliashberg equations we must know how the
six input quantities vary with $x$ because, as shown in ref. 14,
$\lambda_{ij}(x)/\lambda_{ji}(x)=\mu^{*}_{ij}(x)/\mu^{*}_{ji}(x)=N_{N}^{j}(\omega=E_{F},x)/N_{N}^{i}(\omega=E_{F},x)$
where $N_{N}^{j}(\omega=E_{F},x)$ is the density of states at the
Fermi level in the $j$-band and $\lambda_{ij}(x)$ are the
electron-phonon coupling constants. For carrying out this task we
must adopt some drastic approximations because almost all input
quantities are, at the moment, difficult to  determine.

We begin with the four spectral functions of the $MgB_{2}$ that we
can see in Fig. 2 as calculated in ref. 15. We assume that the
general shape of $\alpha^{2}_{ij}F(\omega,x)$ does not change with
$x$ but only the $\lambda_{ij}$ value, that is:
\begin{equation}
\alpha^{2}_{ij}F(\omega,x)=
\frac{\lambda_{ij}(x)}{\lambda_{ij}(x=0)}
\alpha^{2}_{ij}F(\omega,x=0)
\end{equation}
Since it is known that the details of $\alpha^{2}F(\omega)$ do not
affect the resulting $T_c$ significantly \cite{Varelo}, this
approximation is expected to be reasonable.

From the definition of electron-phonon coupling constant we have
\cite{refe18t}:
\begin{equation}
\lambda=\frac{N_{N}(\omega=E_{F})<I^{2}>}{M \Omega^{2}_{0}}
\end{equation}
$M$ is the ion mass, the frequency $\Omega_{0}$ is the frequency
representative of phonon spectrum, $N_{N}$ is the density of
states at the Fermi level and $<I^{2}>$ is the average matrix
element of the electron-ion interaction \cite{refe18t}. The mass M
is the boron mass \cite{Liu} and does not depend on $x$. We assume
that the average matrix element of the electron-ion interaction
$<I^{2}>$ is, in the first approximation, constant because it is
basically determined by the deformation potential which is almost
\cite{Mass} independent of $x$. We have calculated \cite{Mass},
within the virtual crystal approximation, the dependence on $x$ of
$N^{j}_{N}$ (see Fig. 3, upper panel) and we have identified the
representative phonon frequency $\Omega_{0}$ with the $E_{2g}$
phonon mode obtained by experimental data \cite{refe4at} (see Fig.
3 lower panel). So we have that, the more important contribution
to superconductivity in our system, the electron-phonon coupling
constant $\lambda_{\sigma\sigma}$ is:
\begin{equation}
\lambda_{\sigma\sigma}(x)
=\frac{N_{N}^{\sigma}(\omega=E_{F},x)\omega^{2}_{E_{2g}}(x=0)}
{N_{N}^{\sigma}(\omega=E_{F},x=0) \omega^{2}_{E_{2g}}(x)}
\lambda_{\sigma\sigma}(\omega,x=0)
\end{equation}
In this way we assume that the change of $E_{2g}$ influence only
the value of the electron-phonon coupling constant while, in the
first approximation \cite{Varelo}, we neglect its effects on the
shape of the spectral function. For checking this hypothesis we
have also used
$\alpha^{2}_{\sigma\sigma}F_{\sigma\sigma}(\omega,x)=L(\omega,\Omega_{0},\Upsilon)-L(\omega,-\Omega_{0},\Upsilon)$
where $L(\omega,\Omega_{0},\Upsilon)$ is a lorentzian curve with
the peak in $\Omega_{0}(x)=\omega_{E_{2g}}(x)$ and half-width
$\Upsilon(x)$ equal to the error bar of Fig. 2 (lower panel). The
result is almost the same as we can see in Figs. 5 (upper panel)
and 6 (upper panel).

For the other coupling constants, only for simplicity, we assume:
\begin{equation}
\lambda_{ij}(x)
=\frac{N_{N}^{j}(\omega=E_{F},x)}{N_{N}^{j}(\omega=E_{F},x=0)}
\lambda_{ij}(x=0)
\end{equation}
with \cite{refe13t,refe14t} $\lambda_{\sigma\sigma}(x=0)=1.017$,
$\lambda_{\pi\pi}(x=0)=0.448$, $\lambda_{\sigma\pi}(x=0)=0.213$
and $\lambda_{\pi\sigma}(x=0)=0.155$.

In Fig. 4 we can see the calculated electron phonon coupling
constants $\lambda_{ij}$ as a function of $x$. We can see that,
with our approximations, only $\lambda_{\sigma\sigma}$ depends
strongly on x and, for $x>0.33$,
$\lambda_{\sigma\sigma}<\lambda_{\pi\pi}$. The Coulomb
pseudopotential, calculated, for the first time, in ref. 15 and 16
is \cite{refe21t}
\begin{eqnarray}
\hspace{2mm}\mu^{*}(p)= \left| \begin{array}{cc}%
\mu^{*}\ped{\sigma \sigma} & \mu^{*}\ped{\sigma \pi}\\
\mu^{*}\ped{\pi \sigma} & \mu^{*}\ped{\pi \pi}
\end{array} \right| =  \nonumber \\
= \mu(\omega_{c})N^{tot} \ped{N}(E\ped{F},p)
\left| \begin{array}{cc}%
\frac{2.23}{N\ped{N}^{\sigma}(E\ped{F},p)} &
\frac{1}{N\ped{N}^{\sigma}(E\ped{F},p)}\\ & \\
\frac{1}{N\ped{N}^{\pi}(E\ped{F},p)} &
\frac{2.48}{N\ped{N}^{\pi}(E\ped{F},p)}
\end{array} \right| \label{eq:mu}
\end{eqnarray}
where $\mu(\omega_{c},x)$ is a free parameter and and
$N_{N}^{tot}(\omega=E_{F},x)$ is the total normal density of states
at the Fermi level. The numbers 2.23 and 2.48 in the Coulomb matrix
have been calculated in the $MgB_{2}$ case and we suppose that, as
first approximation, they do not depend on $x$ i.e. the Coulomb
pseudopotential depend by $x$ only via the density of states at the
Fermi level while an outer possible dependence by $x$, included the
effect of the disorder, is hidden in $\mu(\omega_{c},x)$ and is the
same for the four values of $\mu^{*}_{ij}(x)$. In conclusion we put
the cut-off energy $\omega_{c}=700$ meV. So we have now only one
free parameter, $\mu(\omega_{c},x)$ that must be determined by
reproducing exactly the experimental critical temperatures.

\begin{figure}[t]
 \begin{center}
 \includegraphics[keepaspectratio, width=0.8\columnwidth]{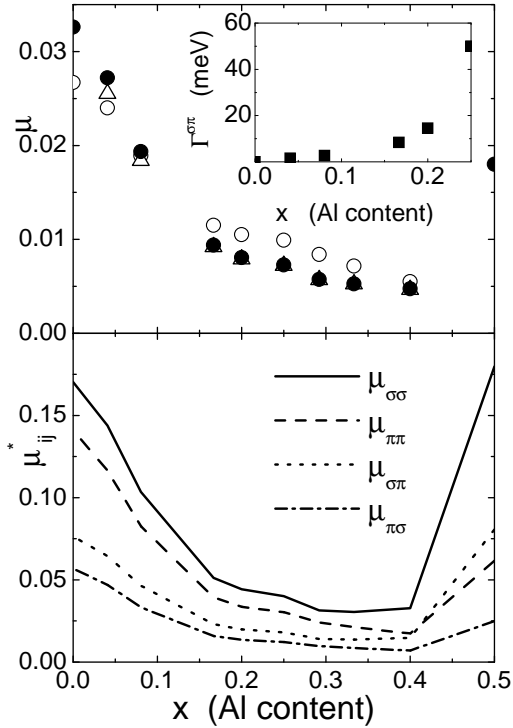}
 \end{center}
  \caption{Upper panel: the prefactor of the Coulomb pseudopotential
 $\mu$ as a function of $x$ when the $\alpha^{2}_{\sigma\sigma}F(\omega,x)$ is a lorentzian (open circles),
 when is $MgB_{2}$-like (filled circles) and when is
 $MgB_{2}$-like and in presence of interband impurities, $\Gamma^{\sigma\pi}=1$ meV, (open up triangles); in the insert the impurity scattering rate
$\Gamma^{\sigma\pi}$, necessary to reproduce
 the experimental $T_{c}$ without change the $MgB_{2}$ physical parameters, versus $x$.
 Lower panel: the values of  Coulomb pseudopotential $\mu^{*}_{ij}$ as a
function of $x$ in the four cases: $\sigma\sigma$ (solid line),
$\pi\pi$ (dashed line), $\sigma\pi$ (dotted line) and $\pi\sigma$
(dashed dotted line) in the
 $MgB_{2}$-like} $\alpha^{2}_{\sigma\sigma}F(\omega,x)$ spectral function case.
 \label{fig:5}
 \end{figure}
 \begin{figure}[t]
 \begin{center}
 \includegraphics[keepaspectratio, width=0.8\columnwidth]{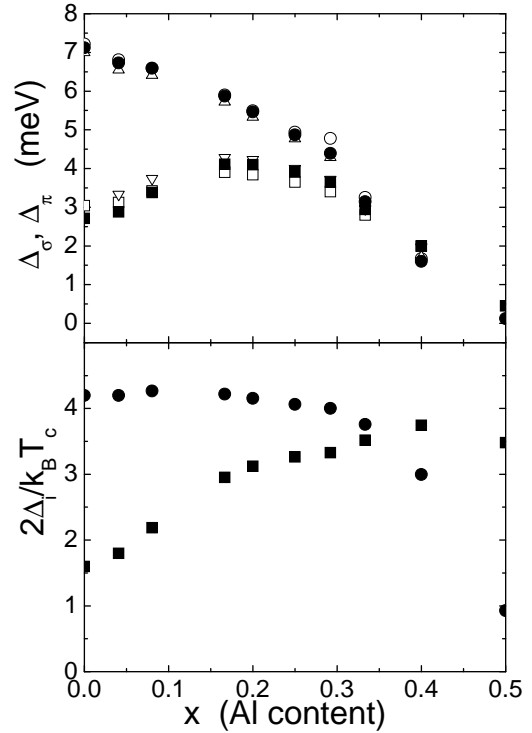}
 \end{center}
  \caption{Upper panel: the value of $\Delta_{i}(i\omega_{n=0})$ of the $\sigma$-band
 (open circles)
 and of the $\pi$-band (open squares) calculated on imaginary axis, at $T=T_{c}/4$, versus x
 when the $\alpha^{2}_{\sigma\sigma}F(\omega,x)$ is a lorentzian,
 when is $MgB_{2}$-like (filled circles and squares) and when is
 $MgB_{2}$-like and in presence of interband impurities, $\Gamma^{\sigma\pi}=1$ meV (open up and down triangles). Lower panel: calculated rate $2\Delta_{i}/k_{B}T_{c}$ for the $\sigma$-band (filled circles)
 and of the $\pi$-band (filled squares) in the
 $MgB_{2}$-like} $\alpha^{2}_{\sigma\sigma}F(\omega,x)$ spectral function case.
 \label{fig:6}
 \end{figure}
 \begin{figure}[t]
 \begin{center}
 \includegraphics[keepaspectratio, width=0.9\columnwidth]{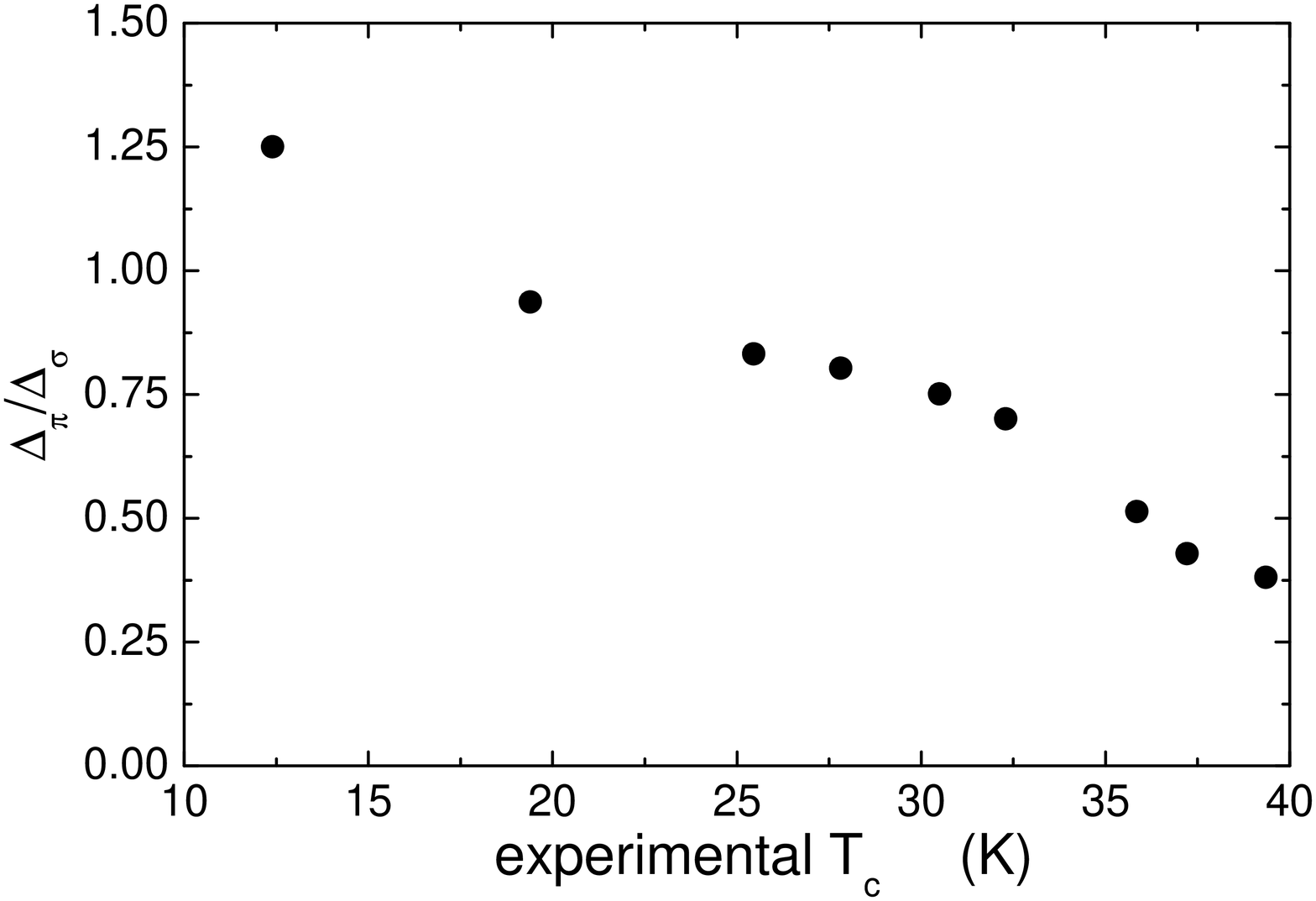}
 \end{center}
  \caption{The calculated rate of
$\Delta_{\pi}/\Delta_{\sigma}$ as a function of experimental
$T_{c}$.} \label{fig:7}
 \end{figure}

In our discussion we have neglected the effects of the disorder
caused by Al doping because we don't know the dependence of
$\Gamma^{ij}$ by $x$ and this fact lead to have a new free
parameter in the problem. We seek of justifying this approximation
and we examine two possibilities. One extreme case is one of
thinking the $Mg_{1-x}Al_{x}B_{2}$ as a contamination with $Al$
impurities of the $MgB_{2}$. In this case also we have only a free
parameter for reproduce the experimental critical temperatures,
$\Gamma^{\sigma\pi}$, because $\Gamma^{\sigma\sigma}$ and
$\Gamma^{\pi\pi}$ don't affect \cite{dirty} $T_{c}$ and
$\Gamma^{\pi\sigma}=N_{N}^{\sigma}(\omega=E_{F})/N_{N}^{\pi}(\omega=E_{F})\Gamma^{\sigma\pi}$.
In this way we find that $\Gamma^{\sigma\pi}$ growths fast in
function of $x$ as we can see in the insert of the upper panel of
Fig. 5. The minimum critical temperature that we can obtain in
this way is $T_{c}\approx 25.8$ K and an isotropic gap
$\Delta_{\sigma}=\Delta_{\pi}\approx 4.1$ meV but we cannot
explain the lower critical temperatures. Moreover the authors of
ref. 27 predict the $\Gamma^{ii}\gg\Gamma^{ij}\approx 1$ meV i.e.
$\Gamma^{ij}$ values found are completely to outside of the range
expected. As second possibility we examine the case \cite{dirty}
$\Gamma^{\sigma\pi}=1$ meV and we find the dependence of
$\mu(\omega_{c})$ by $x$. As we can see in Figs. 5 and 6 (upper
panel) the differences are almost imperceptible. For all these
reasons we neglect the effects of impurities in the Eliashberg
equations.

In the case of $MgB_{2}$ ($x=0$) we find
$\mu(\omega_{c},x=0)=0.0326$ ($\omega_{c}=700$ meV and maximum of
energy equal to 1 eV). In Fig. 5 we show the dependence on $x$ of
$\mu(\omega_{c},x)$ (upper panel) while in the lower panel we can
see the values of $\mu^{*}_{ij}(x)$ calculated following eq. 10. We
note a monotonic parabolic decrease until $x=0.4$ while after it is
present a rapid increase. This result shows that the coulomb
pseudo-potential obtained in this way decreases, approaching the
"shape resonance" determined by the superlattice of boron monolayers
as predicted in several papers \cite{Perali}. The reduction of the
effective Coulomb repulsion in the pairing seems to be a driving
term for rising the critical temperature. The final sharp increase
can be due to an inadequateness of the model: probably, for
$x\approx0.5$, the Migdal's theorem fails \cite{petrus} because
there is a growth of the representative phonon frequencies and a
strong reduction of the Fermi energy.

After determining the dependence of the free parameter
$\mu(\omega_{c})$ on $x$, we can calculate all the physical
quantities by solving the Eliashberg equations. In Fig. 6 we show
the value of $\Delta(i\omega_{n=0})$, at $T=T_{c}/4$, of the
$\sigma$-band and the $\pi$-band (solid circles and solid squares
respectively) versus $x$. The values calculated are almost the
same of those found by solution of the equation
$\Delta_{i}^{0}=\Delta_{i}(\omega=\Delta_{i}^{0})$ on the real
axis with Pad\'{e} approximants because we are in a weak coupling
regime. In strong coupling regime instead the differences can be
remarkable \cite{refe25t}.
 We can see that the behaviour of the $\pi$-gap, as a function of $x$
 has a maximum for $0.164<x<0.198$ while the $\sigma$-gap decreases
 and, for $x\geq0.4$, we find that $\Delta_{\pi}>\Delta_{\sigma}$.
The experimental data \cite{Putti} exist only for $T_{c}\geq 20$ K
($x\leq 0.33$) and are in good agreement with our theoretical
predictions. For higher aluminum content, at present there are no
measurements that confirm or deny the fact that the gaps merge for
$x\simeq0.4$. In the lower panel we show the calculated ratio
 $2\Delta_{i}/k_{B}T_{c}$. Finally in Fig. 7 we can see the calculated rate of
$\Delta_{\pi}/\Delta_{\sigma}$ as a function of experimental
$T_{c}$ (which is preferable due to the loch of agreement between
the different experiments on the true aluminum content of the
sample).

 It is worth to note that if, for the all four
 coupling constants (including the $\sigma\sigma$ component), we use eq. 9, while the
 parameter $\mu(\omega_{c})$ is fixed to the value $\mu(\omega_{c},x=0)=0.0326$ used in the $MgB_{2}$ case,
 i.e. all input quantities of the Eliashberg equations depend by $x$ only via the densities of states at the
 Fermi level, we have a model without free parameters. This model produces a
very good fit of the experimental critical temperatures for
$x<0.33$, but for
 larger $x$ the predicted $T_{c}$ is much lower
 than the experimental values. The value $x=0.33$ corresponds a the transition
 from a pure two-dimensional Fermi surface to an overall three
 dimensional dispersion regime \cite{Mass}. The behavior of the $\sigma$-gap
 is almost unchanged, while the one of
 the $\pi$-gap changes, and decreases monotonically.
 We have no arguments to justify the omission of the
 dependence of $\lambda_{\sigma\sigma}$ on $\omega_{E_{2g}}$.

 Further experimental work is necessary to
 determine the variation of the gaps with $x$ in order
 to understand the correct theoretical approximation and
 to understand the variation of the pairing process
 tuning the Fermi level  through the "shape resonance" condition.

In conclusion we can affirm that not only the $MgB_{2}$ but also
the same material doped with aluminum are weak coupling two band
phononic systems well described by two-bands Eliashberg theory
where the Coulomb pseudopotential and the interchannel pairing
mechanism are key terms to interpret the superconducting phase (it
is the interband pairing that yields the superconductivity in the
$\pi$-band).

This research has been supported by INFM project PRA-UMBRA, INFM
project RSPAIDSCS, INTAS project N. 01-0617 and CNR project
``Progetto 5\% Applicazioni della superconduttivit\`a ad alta
T$_c$".
 \end{document}